

%
%

\def\refto#1{$^{#1}$}

\def\la{\langle}
\def\ra{\rangle}

\def\a{\alpha}
\def\b{\beta}

\def\Tr{{\rm Tr}}

\def\au{{\underline{\alpha}}}
\def\bu{{\underline{\beta}}}



\font\twelverm=cmr10 scaled 1200    \font\twelvei=cmmi10 scaled 1200
\font\twelvesy=cmsy10 scaled 1200   \font\twelveex=cmex10 scaled 1200
\font\twelvebf=cmbx10 scaled 1200   \font\twelvesl=cmsl10 scaled 1200
\font\twelvett=cmtt10 scaled 1200   \font\twelveit=cmti10 scaled 1200
\font\twelvesc=cmcsc10 scaled 1200  
\skewchar\twelvei='177   \skewchar\twelvesy='60


\def\twelvepoint{\normalbaselineskip=12.4pt plus 0.1pt minus 0.1pt
  \abovedisplayskip 12.4pt plus 3pt minus 9pt
  \belowdisplayskip 12.4pt plus 3pt minus 9pt
  \abovedisplayshortskip 0pt plus 3pt
  \belowdisplayshortskip 7.2pt plus 3pt minus 4pt
  \smallskipamount=3.6pt plus1.2pt minus1.2pt
  \medskipamount=7.2pt plus2.4pt minus2.4pt
  \bigskipamount=14.4pt plus4.8pt minus4.8pt
  \def\rm{\fam0\twelverm}          \def\it{\fam\itfam\twelveit}%
  \def\sl{\fam\slfam\twelvesl}     \def\bf{\fam\bffam\twelvebf}%
  \def\mit{\fam 1}                 \def\cal{\fam 2}%
  \def\sc{\twelvesc}               \def\tt{\twelvett}
  \def\sf{\twelvesf}
  \textfont0=\twelverm   \scriptfont0=\tenrm   \scriptscriptfont0=\sevenrm
  \textfont1=\twelvei    \scriptfont1=\teni    \scriptscriptfont1=\seveni
  \textfont2=\twelvesy   \scriptfont2=\tensy   \scriptscriptfont2=\sevensy
  \textfont3=\twelveex   \scriptfont3=\twelveex  \scriptscriptfont3=\twelveex
  \textfont\itfam=\twelveit
  \textfont\slfam=\twelvesl
  \textfont\bffam=\twelvebf \scriptfont\bffam=\tenbf
  \scriptscriptfont\bffam=\sevenbf
  \normalbaselines\rm}



\def\beginlinemode{\endmode
  \begingroup\parskip=0pt \obeylines\def\\{\par}\def\endmode{\par\endgroup}}
\def\beginparmode{\endmode
  \begingroup \def\endmode{\par\endgroup}}
\let\endmode=\par
{\obeylines\gdef\
{}}
\def\singlespace{\baselineskip=\normalbaselineskip}

\def\oneandahalfspace{\baselineskip=\normalbaselineskip
  \multiply\baselineskip by 3 \divide\baselineskip by 2}
\def\doublespace{\baselineskip=\normalbaselineskip \multiply\baselineskip by 2}

\newcount\firstpageno
\firstpageno=2
\footline={\ifnum\pageno<\firstpageno{\hfil}\else{\hfil\twelverm\folio\hfil}\fi}
\def\toppageno{\global\footline={\hfil}\global\headline
  ={\ifnum\pageno<\firstpageno{\hfil}\else{\hfil\twelverm\folio\hfil}\fi}}
\let\rawfootnote=\footnote              
\def\footnote#1#2{{\rm\singlespace\parindent=0pt\parskip=0pt
  \rawfootnote{#1}{#2\hfill\vrule height 0pt depth 6pt width 0pt}}}
\def\raggedcenter{\leftskip=4em plus 12em \rightskip=\leftskip
  \parindent=0pt \parfillskip=0pt \spaceskip=.3333em \xspaceskip=.5em
  \pretolerance=9999 \tolerance=9999
  \hyphenpenalty=9999 \exhyphenpenalty=9999 }
\def\dateline{\rightline{\ifcase\month\or
  January\or February\or March\or April\or May\or June\or
  July\or August\or September\or October\or November\or December\fi
  \space\number\year}}
\def\received{\vskip 3pt plus 0.2fill
 \centerline{\sl (Received\space\ifcase\month\or
  January\or February\or March\or April\or May\or June\or
  July\or August\or September\or October\or November\or December\fi
  \qquad, \number\year)}}


\hsize=6.5truein
\vsize=9.9truein  
\voffset=-1.0truein
\parskip=\medskipamount
\def\\{\cr}
\twelvepoint            
\doublespace            
\overfullrule=0pt       

\def\title                      
  {\null\vskip 3pt plus 0.2fill
   \beginlinemode \doublespace \raggedcenter \bf}

\def\author                     
  {\vskip 3pt plus 0.2fill \beginlinemode
   \singlespace \raggedcenter\sc}

\def\affil                      
  {\vskip 3pt plus 0.1fill \beginlinemode
   \oneandahalfspace \raggedcenter \sl}

\def\abstract                   
  {\vskip 3pt plus 0.3fill \beginparmode
   \singlespace ABSTRACT: }

\def\endtopmatter               
  {\endpage                     
   \body}

\def\body                       
  {\beginparmode}               

\def\head#1{                    
  \goodbreak\vskip 0.5truein    
  {\immediate\write16{#1}
   \raggedcenter \uppercase{#1}\par}
   \nobreak\vskip 0.25truein\nobreak}

\def\beginitems{
\par\medskip\bgroup\def\i##1 {\item{##1}}\def\ii##1 {\itemitem{##1}}
\leftskip=36pt\parskip=0pt}
\def\enditems{\par\egroup}

\def\beneathrel#1\under#2{\mathrel{\mathop{#2}\limits_{#1}}}

\def\refto#1{$^{#1}$}           

\def\references                 
  {\head{References}            
   \beginparmode
   \frenchspacing \parindent=0pt \leftskip=1truecm
   \parskip=8pt plus 3pt \everypar{\hangindent=\parindent}}

\gdef\refis#1{\item{#1.\ }}                     

\gdef\journal#1, #2, #3, 1#4#5#6{               
    {\sl #1~}{\bf #2}, #3 (1#4#5#6)}            

\gdef\refa#1, #2, #3, #4, 1#5#6#7.{\noindent#1, #2 {\bf #3}, #4 (1#5#6#7).\rm}

\gdef\refb#1, #2, #3, #4, 1#5#6#7.{\noindent#1 (1#5#6#7), #2 {\bf #3}, #4.\rm}

\def\pr{\journal Phys.Rev., }

\def\pl{\journal Phys.Lett., }

\def\endreferences{\body}

\def\endpage                    
  {\vfill\eject}

\def\endpaper                   
  {\endmode\vfill\supereject}

\def\ref#1{Ref.~#1}                     
\def\Ref#1{Ref.~#1}                     
\def\[#1]{[\cite{#1}]}
\def\cite#1{{#1}}
\def\(#1){(\call{#1})}
\def\call#1{{#1}}
\def\taghead#1{}
\def\frac#1#2{{#1 \over #2}}

\def\12{{1\over2}}

\catcode`@=11
\newcount\r@fcount \r@fcount=0
\newcount\r@fcurr
\immediate\newwrite\reffile
\newif\ifr@ffile\r@ffilefalse
\def\w@rnwrite#1{\ifr@ffile\immediate\write\reffile{#1}\fi\message{#1}}

\def\writer@f#1>>{}
\def\referencefile{
  \r@ffiletrue\immediate\openout\reffile=\jobname.ref%
  \def\writer@f##1>>{\ifr@ffile\immediate\write\reffile%
    {\noexpand\refis{##1} = \csname r@fnum##1\endcsname = %
     \expandafter\expandafter\expandafter\strip@t\expandafter%
     \meaning\csname r@ftext\csname r@fnum##1\endcsname\endcsname}\fi}%
  \def\strip@t##1>>{}}

\def\citeall#1{\xdef#1##1{#1{\noexpand\cite{##1}}}}
\def\cite#1{\each@rg\citer@nge{#1}}	

\def\each@rg#1#2{{\let\thecsname=#1\expandafter\first@rg#2,\end,}}
\def\first@rg#1,{\thecsname{#1}\apply@rg}	
\def\apply@rg#1,{\ifx\end#1\let\next=\relax
\else,\thecsname{#1}\let\next=\apply@rg\fi\next}

\def\citer@nge#1{\citedor@nge#1-\end-}	
\def\citer@ngeat#1\end-{#1}
\def\citedor@nge#1-#2-{\ifx\end#2\r@featspace#1 
  \else\citel@@p{#1}{#2}\citer@ngeat\fi}	
\def\citel@@p#1#2{\ifnum#1>#2{\errmessage{Reference range #1-#2\space is bad.}%
    \errhelp{If you cite a series of references by the notation M-N, then M and
    N must be integers, and N must be greater than or equal to M.}}\else%
 {\count0=#1\count1=#2\advance\count1 by1\relax\expandafter\r@fcite\the\count0,
  \loop\advance\count0 by1\relax
    \ifnum\count0<\count1,\expandafter\r@fcite\the\count0,%
  \repeat}\fi}

\def\r@featspace#1#2 {\r@fcite#1#2,}	
\def\r@fcite#1,{\ifuncit@d{#1}
    \newr@f{#1}%
    \expandafter\gdef\csname r@ftext\number\r@fcount\endcsname%
                     {\message{Reference #1 to be supplied.}%
                      \writer@f#1>>#1 to be supplied.\par}%
 \fi%
 \csname r@fnum#1\endcsname}
\def\ifuncit@d#1{\expandafter\ifx\csname r@fnum#1\endcsname\relax}%
\def\newr@f#1{\global\advance\r@fcount by1%
    \expandafter\xdef\csname r@fnum#1\endcsname{\number\r@fcount}}

\let\r@fis=\refis			
\def\refis#1#2#3\par{\ifuncit@d{#1}
   \newr@f{#1}%
   \w@rnwrite{Reference #1=\number\r@fcount\space is not cited up to now.}\fi%
  \expandafter\gdef\csname r@ftext\csname r@fnum#1\endcsname\endcsname%
  {\writer@f#1>>#2#3\par}}

\def\ignoreuncited{
   \def\refis##1##2##3\par{\ifuncit@d{##1}%
    \else\expandafter\gdef\csname r@ftext\csname r@fnum##1\endcsname\endcsname%
     {\writer@f##1>>##2##3\par}\fi}}

\def\r@ferr{\endreferences\errmessage{I was expecting to see
\noexpand\endreferences before now;  I have inserted it here.}}
\let\r@ferences=\references
\def\references{\r@ferences\def\endmode{\r@ferr\par\endgroup}}

\let\endr@ferences=\endreferences
\def\endreferences{\r@fcurr=0
  {\loop\ifnum\r@fcurr<\r@fcount
    \advance\r@fcurr by 1\relax\expandafter\r@fis\expandafter{\number\r@fcurr}%
    \csname r@ftext\number\r@fcurr\endcsname%
  \repeat}\gdef\r@ferr{}\endr@ferences}


\let\r@fend=\endpaper\gdef\endpaper{\ifr@ffile
\immediate\write16{Cross References written on []\jobname.REF.}\fi\r@fend}

\catcode`@=12

\citeall\refto		
\citeall\ref		%
\citeall\Ref		%



\hsize=6.0truein
\vsize=8.5truein

\oneandahalfspace


\title{\bf A REVIEW OF THE DECOHERENT HISTORIES
APPROACH TO QUANTUM MECHANICS}
\vskip 0.2in
\author
{J. J. Halliwell}
\affil
{Theory Group, Blackett Laboratory, Imperial College}
{London, SW7 2BZ, United Kingdom}
\abstract{
I review the decoherent (or consistent) histories approach
to quantum mechanics, due to Griffiths, to Gell-Mann and Hartle,
and to Omn\`es. This is an approach to standard quantum theory
specifically designed to apply to genuinely closed systems,
up to and including the entire universe.
It does not depend on an assumed separation of classical and
quantum domains, on notions of measurement, or on collapse of the
wave function. Its primary aim is to find sets of histories for
closed systems exhibiting negligble interference, and therefore,
to which probabilities may be assigned. Such sets of histories
are called consistent or decoherent, and may be manipulated according
to the rules of ordinary (Boolean) logic. The approach provides
a framework from which one may discuss
the emergence of an approximately classical
domain for macroscopic systems, together with
the conventional Copenhagen quantum mechanics for microscropic
subsystems.
In the special case in which the total closed system naturally
separates into a distinguished subsystem coupled to an environment,
the decoherent histories approach is closed related to the quantum
state diffusion approach of Gisin and Percival.
}
\vskip 0.2in
\centerline{\rm (To appear in proceedings of the conference,}
\centerline {\rm { \it Fundamental Problems in Quantum Theory},}
\centerline{\rm Baltimore, June 18-22, 1994, edited by D.Greenberger)}
\vskip 0.2in
\centerline{\rm Imperial College preprint IC/93-94/52. July 1994}
\vskip 0.2in
\body
\leftline{\bf 1. Introduction}
\vglue 0.3cm
Quantum mechanics was originally developed to account for a number
of unexplained phenomena on the atomic scale. The theory
was not thought to be applicable to physics at larger scales,
nor was their felt any need to do so. Indeed, it was
only by reference to  an external, classical, macroscopic world that the
theory could be properly understood. This view of quantum
mechanics, the Copenhagen interpretation, has persisted
for a very long time with not one shred of experimental evidence
against it [\cite{WZ}].

Today, however, more ambitious views of quantum mechanics are
entertained. Experiments have been contemplated ({\it e.g.},
involving SQUIDS) that may probe domains traditionally thought of as
macroscopic [\cite{Leg}]. Even in the absence of such experiments, the
Copenhagen interpretation rests on unsatisfactory foundations.
Macrosopic classical objects are made from microscopic quantum ones.
The dualist view of the Copenhagen interpretation may therefore be
internally inconsistent, and is at best approximate. Most
significantly, there has been a considerable amount of recent
interest in the subject of quantum cosmology in which the notion of
an external classical domain is completely inappropriate [\cite{Hal4}].
Generalizations of conventional quantum theory are required to meet
these new challenges.

John Wheeler was one of the very first people to be so
bold as to even talk about ``the wave function of the universe'' [\cite{Whe}].
He has contributed extensively to our understanding of
quantum mechanics and quantum cosmology, both through his own work,
and through his inspiration of many others in the field.
It is a great pleasure to contribute to this meeting organized in his honour.

\vglue 0.3cm
\noindent{\it 1.1 The Histories Approach}
\vglue 1pt
The object of this paper is to review one particular approach to
quantum mechanics that
was specifically designed to overcome some of the
problems of the orthodox approach.
This is the decoherent (or ``consistent'') histories approch
due to Griffiths [\cite{Gri1,Gri2,Gri3,Gri4,Gri5}],
Gell-Mann and Hartle [\cite{Gel,GH1,GH2,GH3,Har1,Har2,Har3,Har4,Har5,Har6}]
and Omn\`es [\cite{Omn1,Omn2,Omn3,Omn4,Omn5,Omn6,Omn7}].
It is, in particular, a predictive formulation of quantum mechanics
for genuninely closed quantum systems that is sufficiently general
to cope with the needs of quantum cosmology. In brief, its aims are as follows:

\item{\bf 1.} To understand the emergence of an approximately classical
universe from an underlying quantum one, without becoming
embroiled in the details of observers, measuring devices or collapse
of the wave function. Prediction of a classical domain similar to the
one in which we live will generally depend on the initial condition
of the universe, and moreover, could be one of many possibilities
predicted by quantum mechanics. Accommodation, rather
than absolute prediction, of our particular classical universe
may be as much as can be expected.

\item{\bf 2.} To supply a quantum-mechanical framework for reasoning
about the properties of closed physical systems.
Such a framework is necessary if the process of prediction in
quantum mechanics is to be genuinely quantum-mechanical at every
single step. That process
consists of first logically reconstructing
the past history of the universe from records existing in the
classical domain at the present, and then
using the present records together with the
deduced past history
to make predictions about the future (strictly speaking,
about correlations between
records at a fixed moment of time in the future).
A framework for reasoning may also lead to clarification
of many of the conceptually troublesome aspects of quantum mechanics,
such as the EPR paradox.

In more detail, the primary mathematical aim of the histories
approach is to assign probabilities to histories of a closed system.
The approach is a modest generalization of ordinary quantum mechanics, but
relies on a far smaller list of axioms.  These axioms are basically
the statements that the closed system  is described by the usual
mathematical machinery of Hilbert together with a formula for the
probabilities of histories and a rule of interpretation. It makes no
distinction between microscopic and macroscopic, nor does it assume
a ``system-environment'' split; in particular, a separate classical
domain is not assumed. It makes no essential use of
measurement, or collapse of the wave function, although these
notions may be discussed within the framework of the approach. What
replaces measurement is the more general and objective notion of
consistency (or the stronger notion of decoherence), determining
which histories may be assigned probabilities.  The approach also
stresses classical ({\it i.e.} Boolean) logic, the conditions under
which it may be applied, and thus, the conditions under which
ordinary reasoning may be applied to physical system.

The decoherent histories approach is {\it not} designed to answer the
question held by some to be the most important
problem of quantum measurement theory: why
one particular history for the universe ``actually happens''  whilst
the other potential histories allowed by quantum mechanics fade
away.  Although some aspects of this problem are
clarified by the decoherent histories approach,
a satisfactory solution does not appear to be possible
unless something external is added (see Ref.[\cite{Omn8}], for example).
Nor is the approach intended to
meet some philosophical prejudice about the way the world appears to
be.  Its aims are for the large part of a rather pragmatic nature,
namely answering the very physical question of why
the world is described so well by classical mechanics and ordinary
logic, when its atomic  constituents are described by quantum
mechanics.

\vglue 0.3cm
\noindent{\it 1.2 Why histories?}
\vglue 1pt
The basic building blocks in the decoherent histories approach
are the histories of a closed system --
sequences of alternatives at a succession of times.
Why are these objects of particular interest?

\item{(a)} Histories are the most general class of situations one
might be interested in. In a typical experiment, for example,
a particle is emitted from a decaying nucleus at time $t_1$, then it
passes through a magnetic field at time $t_2$, then it is absorbed
by a detector at time $t_3$.

\item{(b)} We would like to understand how classical behaviour can
emerge from the quantum mechanics of closed systems. This involves
showing, amongst other things, that successive positions
in time of a particle, say, are approximately correlated according
to classical laws. This involves the probabilities for
approximate positions at {\it different times}.

\item{(c)} The basic pragmatic aim of theoretical physics is to find
patterns in presently existing data. In cosmology, for example, one
tries to explain the connections between observed data about the
microwave background, the expansion of the universe, the
distribution of matter in the universe, the spectrum of
gravitational waves, {\it etc.}
Why, then, should we not attempt
to formulate our theories in the terms of the density matrix of the
entire universe at the present moment?
There are at least two reasons why not.
First, present records are stored in a wide
variety of different ways -- in computer memories, on photographic
plates, on paper, in our own personal memories, in measuring
devices. The dynamical variables describing those records
could be very hard to identify. The correlations between present records
are far easier to understand in terms of histories.
The patterns in current cosmological
data, for example, are explained most economically by appealing to
the big bang model of the history of the universe.
Second, the correlation between present
records and past events can never be perfect. In order to discuss
the approximate nature of correlations between the past
and the present it becomes necessary to talk about the
histories of a system.

\vglue 0.6cm
\noindent {\bf 2. The Formalism of Decoherent Histories}
\vglue 0.4cm
I now briefly outline the mathematical formalism of the
decoherent histories approach. Further details may be found in
the original papers cited above.
\vglue 0.3cm
\noindent{\it 2.1 Probabilities for Histories}
\vglue 1pt
In quantum mechanics, propositions
about the attributes of a system at a fixed moment of time are
represented by sets of projections operators. The projection
operators $P_{\a}$ effect a partition of the possible alternatives
$\a$ a system may exhibit at each moment of time. They are
exhaustive and exclusive,
$$
\sum_{\a} P_{\a} =1, \quad \quad
P_{\a} P_{\beta} = \delta_{\a \beta} \ P_{\a}
\eqno(2.1)
$$
A projector is said to be {\it fine-grained} if it is of the form
$ | \a \ra \la \a | $, where $\{| \a \ra \}$ are a complete set of
states; otherwise it is {\it coarse-grained}.
A quantum-mechanical history is characterized by a string of
time-dependent projections,
$P_{\a_1}^1(t_1), \cdots P_{\a_n}^n(t_n)$, together with an initial
state $\rho$. The time-dependent projections are related to the
time-independent ones by
$$
P^k_{\a_k}(t_k) = e^{i H(t_k-t_0)} P^k_{\a_k} e^{-i H(t_k-t_0)}
\eqno(2.2)
$$
where $H$ is the Hamiltonian.
The candidate probability for such histories is
$$
p(\a_1, \a_2, \cdots \a_n) = {\rm Tr} \left( P_{\a_n}^n(t_n)\cdots
P_{\a_1}^1(t_1)
\rho P_{\a_1}^1 (t_1) \cdots P_{\a_n}^n (t_n) \right)
\eqno(2.3)
$$
This expression is a familiar one from quantum measurement theory,
but the interpretation is different. Here it is the probability for
a sequence of alternatives for a {\it closed} system. The alternatives at
each moment of time are characterized by projectors.
The projectors are generally not associated with measurements, as they
would be in the Copenhagen view of the formula (2.3). They cannot be
because the system is closed.

It is straightforward to show that (2.3) is both non-negative and
normalized to unity when summed over $\a_1, \cdots \a_n$.
However,
(2.3) does not satisfy all the axioms of probability theory, and for
that reason it is referred to as a candidate probability. It does
not satisfy the requirement
of additivity on disjoint regions of sample space. More precisely,
for each set of histories, one may construct coarser-grained
histories by grouping the histories together. This may be achieved,
for example, by summing over the projections at each moment of time,
$$
{\bar P}_{{\bar \a}} = \sum_{\a \in {\bar \a} } P_{\a}
\eqno(2.4)
$$
(although this is not the most general type of coarse graining).
The additivity requirement is then that the probabilities for each
coarser-grained history should be the sum of the probabilities of
the finer-grained histories of which it is comprised.
Quantum-mechanical interference generally prevents this requirement
from being satisfied; thus histories of closed quantum systems
cannot in general be assigned probabilities.

The standard illustrative example is the double slit experiment. The
histories consist of projections at two moments of time: projections
determining which slit the particle went through at time $t_1$, and
projections determing the point at which the particle hit the screen
at time $t_2$.
As is well-known, the probability distribution for
the interference pattern on the screen cannot be written as a sum of
the probabilities for going through each slit; hence the
candidate probabilities do not satisfy the additivity requirement.

There are, however, certain types of histories for which
interference is negligible, and the candidate probabilities for histories
do satisfy the sum rules.
These histories may be found
using the decoherence functional:
$$
D({\underline {\a}} , {\underline {\a}'} ) =
\Tr \left( P_{\a_n}^n(t_n)\cdots
P_{\a_1}^1(t_1)
\rho P_{\a_1'}^1 (t_1) \cdots P_{\a_n'}^n (t_n) \right)
\eqno(2.5)
$$
Here $ {\underline {\a}} $ denotes the string $\a_1, \a_2, \cdots
\a_n$. Intuitively, the decoherence functional measures the amount
of interference between pairs of histories.
It may be shown that
the additivity requirement is satisfied for all
coarse-grainings if and only if
$$
Re D({\underline {\a}} , {\underline {\a}'} ) = 0
\eqno(2.6)
$$
for all distinct pairs of
histories ${\underline {\a}}, {\underline {\a}'}$ [\cite{Gri1}].
Such sets of histories are said to be {\it consistent}, or
{\it weakly decoherent.} (Note that this definition of consistency is
stronger than that originally introduced by Griffiths [\cite{Gri1}].
See Ref.[\cite{GH2}] for a discussion of this point).

\vglue 0.3cm
\noindent{\it 2.2 Consistency and Classical Logic}
\vglue 1pt
Why are sets of consistent histories are of  interest?
As stated, propositions about the attributes of a quantum system may
be represented by projection operators. The set of all projections
have the mathematical structure of a lattice. This lattice is
non-distributive, and this means that the corresponding propositions
may not be submitted to Boolean logic. Similar remarks hold for
the more complex propositions expressed by general sets of
quantum-mechanical histories.

The reason why {\it consistent} sets of histories are of interest is
that they {\it can} be submitted to Boolean logic. Indeed, a theorem
of Omn\`es states that a set of histories forms a consistent
representation of Boolean logic if and only if it is a consistent
set [\cite{Omn1,Omn6,Omn7}].
That is, in a consistent set of histories, each history
corresponds to a proposition about the properties of a physical
system and we can meaningfully manipulate these propositions without
contradiction using ordinary classical logic. It is in this sense
that the decoherent histories approach supplies a foundation for
reasoning about closed physical systems.

An important example is the case of retrodiction of the past from
present data.
Suppose we have a consistent set of histories.
We would say that the alternative $\a_n$ (present data)
implies the alternatives
$\a_{n-1} \cdots \a_1 $ (past events) if
$$
p(\a_1, \cdots \a_{n-1} | \a_n ) \equiv
{ p(\a_1, \cdots \a_n )  \over p(\a_n) } = 1
\eqno(2.7)
$$
In this way, we can in quantum mechanics build a picture
of the history of the universe, given the present data and the
initial state, using only logic and the consistency of the
histories. We can meaningfully talk about the past properties of
the universe even though there was no measuring device there to
record them.

There is, however, a caveat. It is very frequently the case that the
same initial state and present data will admit two or more
inequivalent sets of consistent  histories the union of which is not
a consistent set. There then often exist propositions about the past
properties of the system that are  logically implied by the present
data in some sets of histories but not in others.  Omn\`es refers to
such propositions as ``reliable'', whilst propositions that are
implied by the present data in every consistent set of histories are
labeled ``true'' [\cite{Omn9}] (see also Ref.[\cite{Esp}]).
The existence of these so-called multiple logics
means that one cannot say that past properties corresponding
to reliable propositions ``actually
happened'', because they depend on a particular choice of
consistent histories. In the histories approach, the reconstruction
of history from present records is therefore {\it not unique}. This
means that the approach does not in general allow one to talk about
the past history of the universe ``the way it really is''.

Is this a problem? Some feel that it is [\cite{DK}].
For the immediate practical purposes of quantum cosmology, however,
it does not appear to be a difficulty. Recall that what
quantum mechanics must ultimately explain is the correlation between
records at a fixed moment of time. As stated earlier, it is easiest
to understand those correlations in terms of histories, but
histories enter as an intermediate step. The correlations
between two records at a fixed moment of time predicted by quantum
mechanics are unambiguous, even though the histories corresponding
to these records may not be unique.

\vglue 0.6cm
\noindent {\bf 3. Decoherence, Correlation and Records}
\vglue 0.4cm
How may the consistency condition (2.6)
come to be satisfied? First of all, it is
straightforward to show that, with some exceptions,
histories of completely fine-grained
projection operators will generally {\it not} lead to consistency.
The consistency condition is generally satisfied only by
sets of histories that are {\it coarse-grained}.
When sets of histories satisfy the consistency
condition (2.6) as a result of coarse-graining,
they typically satisfy, in addition, the stronger condition that
both the real and imaginary parts of the off-diagonal terms of the
decoherence functional vanish,
$$
D(\au, \au') = 0, \quad for \quad \au \ne \au'
\eqno(3.1)
$$
This I shall refer to quite simply as {\it decoherence}.
(It is sometimes referred to more specifically as
{\it medium} decoherence [\cite{GH2}] but we shall not do so here).

Physically, decoherence
is intimately related the existence of records
about the system somewhere in the universe.
In this sense decoherence replaces and generalizes the
notion of measurement in ordinary quantum mechanics.
Sets of histories decohere, and hence the system ``acquires
definite properties'', not necessarily through measurement, but
through the interactions and correlations of the variables that are
followed with the variables that are ignored as a result of
the coarse-graining.

Decoherence is typically only approximate so measures of approximate
decoherence are required. First, note that
the decoherence functional obeys the simple inequality [\cite{DH}],
$$
\bigl| D(\au, \au') \bigr|^2  \ \le \ D(\au,\au) \ D(\au', \au')
\eqno(3.2)
$$
Intuitively, this result indicates that there can be no interference
with a history which has candidate probability zero.
It also suggests a possible measure of approximate decoherence:
we say that
a system decoheres to order $\epsilon$ if the decoherence functional
satisfies (3.2) with a factor of $\epsilon^2$ multiplying
right-hand side. This condition may be shown to imply that
most (but not all) probability sum rules will then be satisfied to
order $\epsilon$ [\cite{DH}].

Approximate decoherence to order $\epsilon$
means that the probabilities are defined only up to that
order. In typical cases, $\epsilon$ is substantially smaller than any
other effect that could conceivably modify the probabilities,
and hence they may be thought of as precisely
defined for all practical purposes. Alternatively, it has been
conjectured that a generic approximately decoherent set of histories
may be turned into an exactly decoherent set by modifying to
order $\epsilon$ the
operators projected onto at each moment of time [\cite{DK}].

\vglue 0.3cm
\noindent {\it 3.1 Records Imply Decoherence}
\vglue 1pt
I now exemplify the connection between
records and decoherence.
Consider a closed system $S$ which consists of two weakly
interacting
subsystems $A$ and $B$. The Hilbert space ${\cal H}$ of $S$ is
therefore of the form ${\cal H}_A \otimes {\cal H}_B$.
For simplicity let ${\cal H}_A $ and ${\cal H}_B$ have the same
dimension.
Suppose we are interested in the histories characterized solely by
properties of system $A$, thus $B$ is regarded as the environment.
The system is analyzed using the decoherence functional (2.5), where
we take the $P_{\a}$ to denote a projection on ${\cal H}_A$ (
hence the projections in the decoherence functional are of the form
$P_{\a} \otimes I^B$, where $I^B$ denotes
the identity on ${\cal H}_B$).
I also introduce projections $R_{\b}$ on the Hilbert space ${\cal H}_B$.

I shall show that histories of $A$ satisfy the decoherence
condition (3.1) if the sequences
of alternatives the histories consist of exhibit {\it exact} and
{\it persistent} correlations with sequences of alternatives of $B$.
To be precise,
suppose that the alternatives of $A$ characterized by
$P_{\a_k}^k$ at each moment of time $t_k$ are perfectly recorded in
$B$ as a result of their interaction. Suppose also that this
record in $B$ is perfectly persistent ({\it i.e.}, permanent).
This means that at any time $t_f$ after
the time $t_n$ of the last projection on $A$ there exist a sequence
of alternatives of $B$, $\b_1, \cdots \b_n$, that are in perfect
correlation with the alternatives of $A$, $\a_1 \cdots \a_n$ at
times $t_1 \cdots t_n $.

For each moment of time $t_k$,
the decoherence functional (2.5) may be written,
$$
D(\au, \au') = \sum_{\b_k} \ \Tr \left(
I^A \otimes R_{\b_k}^k \cdots P_{\a_k}^k \otimes I^B \cdots
\ \rho \ \cdots P_{\a_k'}^k \otimes I^B \cdots \right)
\eqno(3.4)
$$
using the exhaustivity of the projections $R_{\b_k}^k$,
where the dots denote the projections at times other than $t_k$ and
the unitary evolution operators between them.
Now, since $R_{\b_k}^k$ a projector, it may be replaced by
$(R_{\b_k}^k)^2$. Furthermore, the assumption of persistence then allows
us to move the projector
$R_{\b_k}^k$ through all the unitary evolution operators occuring
after time $t_k$ on each side of the decoherence functional, with
the result,
$$
D(\au, \au') = \sum_{\b_k} \ \Tr \left(
\cdots P_{\a_k}^k \otimes R_{\b_k}^k \cdots
\ \rho \
\cdots P_{\a_k'}^k \otimes R_{\b_k}^k \cdots
\right)
\eqno(3.5)
$$
Finally, the assumed correlated between the
alternative
$\a_k$ in $A$ and $\b_k$ in $B$ means that the terms of the form
$P_{\a_k}^k \otimes R_{\b_k}^k$ on each side will yield zero when operating
on everything that came earlier in the chain, unless $\a_k = \b_k$.
Eq.(3.5) will therefore be diagonal in $\a_k$.
Repeating the argument for all other values of $k$,
we thus find that, as advertized,
a perfect and persistent correlation of
alternatives of $A$ with those of $B$ leads to exact decoherence of
the histories of $A$.
It is not just the consistency condition (2.6) that is satisfied
through persistent correlation with another
subsystem, but the stronger condition of decoherence, (3.1).
This argument was inspired by an argument given by Hartle
[\cite{Har1}] in his
discussion of the recovery of the Copenhagen interpretation from the
decoherent histories approach. A more detailed version of it is
given in Ref.[\cite{Hal1}].

\vglue 0.3cm
\noindent{\it 3.2 Decoherence Implies Generalized Records}
\vglue 1pt
There is a converse to the above result, namely that
Eq.(3.1), in a certain sense, implies
the existence of records [\cite{GH2}].
Consider the decoherence functional (2.5),
for any system (not just the special one discussed above).
Introduce the convenient notation
$$
C_{\au} = P_{\a_n}(t_n) \cdots P_{\a_1}(t_1)
\eqno(3.9)
$$
Let the initial state be pure, $\rho = | \Psi \ra \la \Psi |$.
In this case, the decoherence condition (3.1) is referred to as
{\it medium decoherence}.
It implies that the states $ C_{\au} | \Psi \ra $
are an orthogonal (but in general incomplete) set. There therefore
exists a set of projection operators $R_{\bu}$ (not in general unique)
of which these states are eigenstates,
$$
R_{\bu} \ C_{\au} | \Psi \ra = \delta_{\au \bu} \ C_{\au} | \Psi \ra
\eqno(3.10)
$$
Note that the $C_{\au}$'s are not themselves projectors in general.
One may then consider histories consisting of the string of
projections (3.9), adjoined by the projections $R_{\bu}$ at any time
after the final time. The decoherence functional for such histories
is
$$
D(\au, \bu | \au', \bu' ) = \Tr \left( R_{\bu} C_{\au} | \Psi \ra
\la \Psi | C^{\dag}_{\au'}  R_{\bu'} \right)
\eqno(3.11)
$$
These extended histories decohere exactly by virtue of (3.10) and (3.1),
and thus the
diagonal elements of (3.11), which we denote $p(\au, \bu)$, are true
probabilities. The correlations contained in these probabilities
may therefore be discussed. Indeed, Eq.(3.10) implies that
$p(\au,\bu) = \delta_{\au \bu} \ p(\au) $, and thus $\au$ and $\bu$
are perfectly correlated.

Medium decoherence therefore implies the existence of a string of
alternatives $\b_1 \cdots \b_n$, at some fixed moment of time after $t_n$,
perfectly correlated with the string
$\a_1, \cdots \a_n$ at the sequence of times $t_1 \cdots t_n$.
For this reason the projection operators
$R_{\au}$ are referred to as generalized records: information about
the histories characterized by alternatives  $\a_1 \cdots \a_n$ is
recorded somewhere. It is, however, not possible to say that the
information resides in a particular subsystem, since we have not
specified the form of the system $S$; indeed, it is generally not
possible to divide it into subsystems.

\vglue 0.6cm
\noindent {\bf 4. Towards a Quasiclassical Domain}
\vglue 0.4cm
Given the framework sketched above, one of the principle aims of
the decoherent histories approach is to demonstrate the emergence of
an approximately classical world from an underlying quantum one,
together with the quantum fluctuations about it described by the
familiar Copenhagen quantum mechanics of measured subsystems.
Such a state of affairs is referred to as a quasiclassical domain
[\cite{GH1,GH2,GH3}].
In more technical terms, a quasiclassical domain
consists of a decoherent set of
histories, characterized largely by the same types of variables at
different times, and whose probabilities are peaked about
deterministic evolution equations for the variables characterizing
the histories.

The histories should, moreover, be {\it maximally refined} with
respect to a specified degree of approximate decoherence.  That is,
one specifies a decoherence factor $\epsilon$ in the approximate
decoherence condition discussed above. This should, for example, be
chosen so that the probabilities are defined to a precision far
beyond any conceivable test. Then, the histories should be fine-grained
({\it e.g.}, by reducing the widths of the projections) to the point
that further fine-graining would lead to violation of the
specified degree of approximate decoherence. The resulting set of
histories are then called maximally refined. The reason for
maximally refining the histories is to reduce as much as possible
any apparent subjective element in the choice of coarse-graining.

Given the Hamiltonian and initial state of the system, one's task is
to compute the decoherence functional for various different choices
of histories, and see which ones lead to quasiclassical behaviour.
As suggested by the discussion at the end of Section 2,
there could be -- and probably are -- many such sets
of variables leading to quasiclassical behaviour. An important
problem is to find as many such sets as possible and develop
criteria to distinguish between them. One useful criterion is
whether a quasiclassical domain can support the existence of an
information gathering and utilizing system, or IGUS. This is a
complex adaptive system that exploits the regularities in its
environment in such a way as to ensure its own survival.
This particular criterion may rule out domains described by
particularly bizarre decoherent sets of histories, such as ones
described by completely different variables at each moment of time,
because the IGUS may not have sufficient information processing
capabilities to assimilate its environment.
Also, criteria such as the existence of IGUSes alleviate to some
degree the multiplicity of consistent sets of histories discussed in
Section 2.
These issues are discussed further in
Refs.[\cite{GH1,GH2,GH3,DK,Sau1,Sau2,Sau3}]

\vglue 0.3cm
\noindent{\it 4.1 Histories of Hydrodynamic Variables}
\vglue 1pt
What are the sets of variables that can lead to quasiclassical
behaviour?
One particular set of variables that are strong candidates for
it are the integrals over small volumes of
locally conserved densities. A generic system will usually not have
a natural separation into ``system'' and ``environment'', and it is
one of the strengths of the decoherent histories approach that it
does not rely on such a separation.
Certain variables will, however, be distinguished by the existence
conservation laws for total energy, momentum, charge, particle
number, {\it etc.} Associated with such conservation laws are local
conservation laws of the form
$$
{\partial \rho \over \partial t} +{\bf  \nabla} \cdot {\bf j} = 0
\eqno(4.1)
$$
The candidate quasiclassical variables are then
$$
Q_V = \int_{V} d^3 x \ \rho({\bf x})
\eqno(4.2)
$$
If the volume $V$ over which the local densities are smeared is
infinite, $Q_V$ will be an exactly conserved quantity. In quantum
mechanics it will commute with the Hamiltonian, and, as is easily
seen, histories of $Q_V$'s will decohere exactly.
If the volume is finite but large compared to
the microscopic scale, $Q_V$ will be slowly varying compared to all
other dynamical variables. This is because the local conservation
law (4.1) permits $Q_V$ to change only by {\it redistribution}, which
is limited by the rate at which the locally conserved quantity can
flow out of the volume. Because these quantities are slowly varying,
histories of them should approximately decohere.
Furthermore, the fact that the $Q_V$'s are slowly varying may also be
used, at least classically, to derive an approximately closed set of
equations involving only those quantities singled out by the
conservation laws. These equations are, for example, the
Navier-Stokes equations, and the derivation of them is a standard
(although generally non-trivial) exercise in non-equilibrium
statistical mechanics [\cite{For}]. One of the current goals of the decoherent
histories approach is to reexpress this derivativion in the language
of histories [\cite{HH}].

\vglue 0.3cm
\noindent{\it 4.2 Quantum Brownian Motion Models}
\vglue 1pt
Many concrete investigations of the mechanics of decoherence have
actually concerned quantum Brownian motion models, primarily
because calculations can be carried out with comparative ease
[\cite{GH2,DH}].
These have proved to be quite instructive.
Very briefly, such models
consist of a particle of mass $M$ in a potential $V(x)$
linearly coupled to an environment consisting of a large
bath of harmonic oscillators in a thermal state at temperature $T$,
and characterized by a dissipation coefficient $\gamma$.
The types of histories commonly considered are sequences of
approximate positions of the Brownian particle,
specified up to some width $\sigma$, whilst the environment of
oscillators is traced over.

The results may briefly be summarized as
follows. Decoherence through interaction with the environment
is an extremely effective process. For example, for a particle whose
macroscopic parameters (mass, timescale, {\it etc.}) are of order $1$
in c.g.s. units,  and for an environment at room temperature,
the degree of approximate decoherence is of order
$ \exp\left( - 10^{40} \right)$, a {\it very} small number.
The probabilities for histories of positions are then
strongly peaked about the classical equations of motion,
but modified by a dissipation term,
$$
M \ddot x + M \gamma \dot x + V'(x) =0
\eqno(4.3)
$$
There are fluctuations about classical predictability, consisting of
the ubiquitous quantum fluctuations, adjoined by thermal
fluctuations from the interaction with the environment.
There is a generally a tension between the
demands of decoherence and classical predictability,
due to the fact that
the degree of decoherence improves with increasing environment
temperature, but predictability deteriorates, because the
fluctuations about (4.3) grow.
However, if the particle is sufficiently massive, it can resist the
thermal fluctuations and a compromise regime can be found in which
there is a reasonable degree of both decoherence and classical predictability.

\vglue 0.6cm
\noindent{\bf 5. Decoherent Histories and Quantum State Diffusion}
\vglue 0.4cm
The decoherent histories approach is closely connected to the
quantum state diffusion (QSD) approach to open systems.
In that approach, the master equation for the reduced density
operator of an open system (essentially a closed system in which one
focuses on a particular subsystem) is solved by
exploiting a purely mathematical connection
with a certain non-linear stochastic Schr\"odinger equation (Ito
equation) [\cite{GP}].
Solutions to the Ito equation turn out to correspond rather closely
to the results of actual laboratory experiments ({\it e.g.}, in
quantum optics), and are therefore held to describe
individual systems and processes. For example, in a quantum Brownian
motion model, the solutions to the Ito equation become localized
about points in phase space following the classical equations of
motion. The connection with the decoherent histories approach is
that, loosely speaking, the solutions to the Ito equation may be
thought of as the individual histories belonging to a decoherent
set [\cite{Dio5}]. More precisely,
the variables that localize in the QSD approach also define a
decoherent set of histories in the decoherent histories approach.
The degrees of localization and of decoherence are related, and the
probabilities assigned to histories in each case are essentially the
same. This connection could be a very useful one, both conceptually
and computationally, and efforts to exploit it are being made.

\vglue 0.6cm
\noindent{\bf 6. What Have We Gained?}
\vglue 0.4cm
In this contribution I have tried to give a brief overview
of the decoherent histories approach to quantum
theory. What has the decoherent histories approach taught us?

At the level of ordinary quantum mechanics, applied to laboratory
situations, two things have been gained. First of all, a minimal
view of the decoherent histories approach is that it is
in a sense a more
refined version of the Copenhagen interpretation.  It rests on a
considerable smaller number of axioms, and in particular, it is a
predictive formulation of quantum mechanics that does not rely on
any kind of assumptions referring to measurement or to a classical
domain. It is internally consistent and reproduces all the
experimental predictions of the Copenhagen approach.
Secondly, it provides a clear set of criteria for the
application of ordinary logic in quantum mechanics. Since many of
the conceptual difficulties of quantum mechanics are essentially
logical ones, {\it e.g.}, the EPR paradox, a clarification of the
applicability
of logic has been argued to lead to
their resolution [\cite{Gri3,Omn2,Omn5}].
Such a resolution is not strictly possible in Copenhagen
quantum mechanics, because it does not offer clear guidelines
for the application of ordinary logic.

There will, of course, always be some who claim that they can finesse
their way through any difficulty of quantum mechanics without having
to worry about the somewhat cumbersome machinery of the histories
approach described here. In this connection, Omn\`es has to say the
following [\cite{Omn7}]:

\item{} ``It may be true, as some people say, that everything is in
Bohr, but this has been a matter for hermeneutics, with the endless
disputes any scripture will lead to. It may also happen that he
guessed the right answers, but the pedagogical means and the
necessary technique details were not yet available to him. Science
cannot, however, proceed by quotations, however elevated the source.
It proceeds by elucidation, so that feats of genius can become
ordinary learning for beginners.''

Intuition alone may be sufficient to see some through the
difficulties of non-relativistic quantum mechanics, but if we are to
extend quantum theory to the entire universe, a reliable vehicle
for travel beyond the domain of our intuition is required.
For quantum cosmology, the development of the decoherent
histories approach has been a considerable bonus.
The decoherent histories
approach supplies an unambiguous, workable and predictive scheme
for actually applying quantum mechanics to genuinely closed systems.
Furthermore, as discussed at
some length in this paper, it supplies a conceptually clear method
of discussing the emergence of classicality in closed quantum
systems, and this is perhaps its greatest success.

Still outstanding are the largely technical difficulties of
quantum cosmology connected with quantizing gravity. However, it is
possible that the histories approach might be of use there also. The
focus on histories may circumvent the ``problem of time''
encountered in most canonical approaches to quantum gravity.
Isham and collaborators,
for example, are currently exploring the possibility of
histories-based formulations of quantum theory that
do not rely on the conventional Hilbert space structure, or on the
existence of a preferred time coordinate [\cite{Ish,IL,ILS}],
building on an earlier
suggestion of Hartle [\cite{Har3}]. Much remains to be done, but on both
conceptual and technical grounds, the histories approach to quantum
cosmology appears to be a particularly promising avenue for future
research.

Further aspects of the decoherent histories approach are discussed
in \break
Refs.[\cite{Alb1,Alb2,Ble1,Bru1,Bru2,CH,Dio1,Dio2,Dio3,Dio4,Fin,GPa,Hal2,Hal3,
LM,Ohk,PZ,Sor,Twa1,Twa2,Zur}]

\vglue 0.6cm
\noindent{\bf 7. Acknowledgments}
\vglue 0.4cm
I am grateful to the organizers for giving me the opportunity to
take part in such an interesting meeting.
I would also like to thank numerous colleagues for useful
conversations,
especially Lajos Di\'osi,
Fay Dowker, Murray Gell-Mann, Nicolas Gisin,
Jim Hartle, Chris Isham, Adrian Kent, Seth Lloyd,
Roland Omn\`es, Ian Percival,
Trevor Samols, Dieter Zeh and Wojciech Zurek.
This work was supported by a University Research Fellowship from the
Royal Society.

\vglue 0.6cm
\noindent{\bf 8. References \hfil}
\vglue 0.4cm

\medskip

\def\pr{{\it Phys.Rev.\ }}

\def\pl{{\it Phys.Lett.\ }}

\refis{Alb1} A. Albrecht, {\sl Phys.Rev.} {\bf D46}, 5504 (1092).
{\it Investigating Decoherence in a Simple System.}

\refis{Alb2} A. Albrecht, Imperial Preprint 92-93/03 (1992).
{\it Following a Collapsing Wave Function.}


\refis{Ble1} M. Blencowe, {\sl Ann.Phys.} {\bf 211}, 87 (1991).
{\it The Consistent Histories
Interpretation of Quantum Fields in Curved Spacetime.}

\refis{Bru1} T. Brun, {\sl Phys. Rev.} {\bf D47}, 3383 (1993).
{\it Quasiclassical Equations of Motion for Nonlinear Brownian
Systems.}

\refis{Bru2} T. Brun, Caltech preprint (1994).
{\it The Decoherence of Phase Space Histories}.

\refis{CH} E. Calzetta and B. L. Hu, in {\it Directions in General
Relativity}, edited by B. L. Hu and T. A. Jacobson (Cambridge
University Press, Cambridge, 1993).
{\it Decoherence of Correlation Histories.}

\refis{Dio1} L. Di\'osi, Budapest preprint KFKI-1992-03/A (1992).
{\it Coarse Graining and Decoherence Translated into von Neumann
Language.}

\refis{Dio2} L. Di\'osi, Budapest preprint KFKI-RMKI-23 (1993).
{\it Unique Family of Consistent Histories in the Caldeira-Leggett
Model.}

\refis{Dio3} L. Di\'osi, Budapest preprint KFKI-RMKI-25 (1993).
{\it Comment on, ``Consistent Intepretation of Quantum Mechanics
Using Quantum Trajectories''.}

\refis{Dio4} L. Di\'osi, Budapest preprint KFKI-RMKI-28 (1993).
{\it Unique Quantum Paths by Continuous Diagonalization of the
Density Operator.}

\refis{Dio5} L. Di\'osi, N. Gisin, J. J. Halliwell and I.C.Percival,
Imperial College preprint IC 93-94/25 (1994).
{\it Decoherent Histories and Quantum State Diffusion}.

\refis{DK} H. F. Dowker and A. Kent, Newton Institute preprint NI-94006
(1994).
{\it On the Consistent Histories Formulation of Quantum Mechanics}.

\refis{DH} H. F. Dowker and J. J. Halliwell, {\sl Phys. Rev.} {\bf
D46}, 1580 (1992).
{\it Quantum Mechanics of History: The Decoherence Functional in
Quantum Mechanics.}

\refis{Esp} B. d'Espagnat, {\sl J.Stat.Phys.} {\bf 56}, 747 (1989).
{\it Are There Realistically Interpretable Local Theories?}

\refis{Fin} J. Finkelstein, San Jos\'e preprint SJSU/TP-93-10 (1993).
{\it On the Definition of Decoherence.}

\refis{For} D. Forster, {\it Hydrodynamic Fluctuations, Broken
Symmetry and Correlation Functions} (Benjamin, Reading, MA, 1975).

\refis{Gel} M. Gell-Mann, {\it The Quark and the Jaguar}
(Little, Brown and Company, London, 1994).

\refis{GH1} M. Gell-Mann and J. B. Hartle, in {\it Complexity, Entropy
and the Physics of Information, SFI Studies in the Sciences of Complexity},
Vol. VIII, W. Zurek (ed.) (Addison Wesley, Reading, 1990); and in
{\it Proceedings of the Third International Symposium on the Foundations of
Quantum Mechanics in the Light of New Technology}, S. Kobayashi, H. Ezawa,
Y. Murayama and S. Nomura (eds.) (Physical Society of Japan, Tokyo, 1990).
{\it Quantum Mechanics in the Light of Quantum Cosmology.}

\refis{GH2} M. Gell-Mann and J. B. Hartle, {\sl Phys.Rev.} {\bf D47},
3345 (1993).
{\it Classical Equations for Quantum Systems}.

\refis{GH3} M. Gell-Mann and J. B. Hartle, Santa Barbara preprint
(1994).
{\it Equivalent Sets of Histories and Multiple Quasiclassical Domains.}

\refis{GP} N. Gisin and I.C. Percival, {\sl J.Phys.} {\bf A25},
5677 (1992). {\it The Quantum State Diffusion Model Applied to Open
Systems}.

\refis{GPa} S. Goldstein and D. N. Page, University of Alberta preprint
Alberta-Thy-43-93, gr-qc 9403055 (1993).
{\it Linearly Positive Histories.}

\refis{Gri1} R. B. Griffiths, {\sl J.Stat.Phys.} {\bf 36}, 219 (1984).
{\it Consistent Histories and the Interpretation of Quantum Mechanics}.

\refis{Gri2} R. B. Griffiths, {\sl Phys.Rev.Lett.} {\bf 70}, 2201
(1993).
{\it Consistent Interpretations of Quantum Mechanics Using Quantum
Trajectories.}

\refis{Gri3} R. B. Griffiths, {\sl Am.J.Phys.} {\bf 55}, 11 (1987).
{\it Correlations in Separated Quantum Systems: A Consistent
History Analysis of the EPR Problem.}

\refis{Gri4} R. B. Griffiths, Pittsburgh preprint (1994).
{\it A Consistent History Approach to the Logic of Quantum
Mechanics.}

\refis{Gri5} R. Griffiths, Pittsburgh preprint (1994).
{\it Consistent Quantum Counterfactuals}.

\refis{Hal1} J. J. Halliwell, in
{\it Stochastic Evolution of Quantum States in Open Systems and
Measurement Processes}, edited by L. Di\'osi (World Scientifc,
Singapore, 1994).
{\it Aspects of the Decoherent Histories Approach to Quantum Mechanics}

\refis{Hal2} J. J. Halliwell, \pr {\bf D48}, 2739 (1993).
{\it Quantum Mechanical Histories and the Uncertainty Principle:
Information-Theoretic Inequalities}.

\refis{Hal3} J. J. Halliwell, \pr {\bf D48}, 4785 (1993).
{\it Quantum Mechanical Histories and the Uncertainty Principle: II.
Fluctuations about Classical Predictability}.

\refis{Hal4} J. J. Halliwell, in {\it General Relativity and Gravitation
1992}, edited by R. J. Gleiser, C. N. Kozameh and O. M. Moreschi
(IOP Publishers, Bristol, 1993).
{\it The Interpretation of Quantum Cosmological Models}.

\refis{HH} J. J. Halliwell and J. B. Hartle, work in progress.

\refis{Har1} J. B. Hartle, in {\it Quantum Cosmology and Baby
Universes}, S. Coleman, J. Hartle, T. Piran and S. Weinberg (eds.)
(World Scientific, Singapore, 1991).  {\it The Quantum Mechanics of
Cosmology.}

\refis{Har2} J. B. Hartle, {\sl Phys.Rev.} {\bf D44}, 3173 (1991).
{\it Spacetime Coarse-Grainings in Non-Relativistic Quantum Mechanics.}

\refis{Har3} J. B. Hartle, in, Proceedings of the 1992 Les Houches Summer
School, {\it Gravitation et Quantifications}, B.Julia (ed.).
{\it Spacetime Quantum Mechanics and the Quantum Mechanics of Spacetime.}

\refis{Har4} J. B. Hartle, in the festschrift for C. Misner, edited by
B.L. Hu, M. P. Ryan amd C. V. Vishveshwara (Cambridge University
Press, Cambridge, 1993).
{\it The Quantum Mechanics of Closed Systems.}

\refis{Har5} J. B. Hartle, in the festschrift for D. Brill, edited by
B.L. Hu and T. Jacobson (Cambridge University Press, Cambridge, 1993).
{\it The Reduction of the State Vector and Limitations on
Measurement in the Quantum Mechanics of Closed Systems.}

\refis{Har6} J. B. Hartle, Santa Barbara preprint (1994), to appear
in {\it Proceedings of the Lanczos Centenary Meeting}.
{\it Quasiclassical Domains in a Quantum Universe}.

\refis{Ish} C. Isham, Imperial Preprint IC/92-93/39 (1993).
{\it Quantum Logic and the Histories Approach to Quantum Theory.}

\refis{IL} C. Isham and N. Linden, Imperial College preprint
IC/TP/93-94/35 (1994).
{\it Quantum Temporal Logic and Decoherence Functionals in the
Histories Approach to Generalized Quantum Theory}.

\refis{ILS} C. Isham, N. Linden and S.Schreckenberg,
Imperial College preprint IC/TP/93-94/42 (1994).
{\it The Classification of Decoherence Functionals:
An Analogue of Gleason's Theorem}.

\refis{LM} R. Laflamme and A. Matacz, Los Alamos preprint (1993).
{\it Decoherence Functional and Inhomogeneities in the Early
Universe.}

\refis{Leg} A. J. Leggett, {\sl Suppl.Prog.Theor.Phys.} {\bf 69}, 80 (1980).
{\it Macroscopic Quantum Systems and the Quantum Theory of Measurement.}

\refis{Ohk} Y. Ohkuwa, Santa Barbara preprint, UCSBTH-92-40 (1992).
{\it Decoherence Functional and Probability Interpretation.}

\refis{Omn1} R. Omn\`es, {\sl J.Stat.Phys.} {\bf 53}, 893 (1988).
{\it Logical Reformulation of Quantum Mechanics. I. Foundations.}

\refis{Omn2} R. Omn\`es, {\sl J.Stat.Phys.} {\bf 53}, 933 (1988).
{\it Logical Reformulation of Quantum Mechanics. II.
Interferences and the Einstein-Podolsky-Rosen Experiment.}

\refis{Omn3} R. Omn\`es, {\sl J.Stat.Phys.} {\bf 53}, 957 (1988).
{\it Logical Reformulation of Quantum Mechanics. III.
Classical Limit and Irreversibility.}

\refis{Omn4} R. Omn\`es, {\sl J.Stat.Phys.} {\bf 57}, 357 (1989).
{\it Logical Reformulation of Quantum Mechanics. IV.
Projectors in Semiclassical Physics.}

\refis{Omn5} R. Omn\`es, {\sl Phys.Lett.} {\bf A138}, 157 (1989).
{\it The Einstein-Podolsky-Rosen Problem: A New Solution.}

\refis{Omn6} R. Omn\`es, {\sl Ann.Phys.} {\bf 201}, 354 (1990).
{\it From Hilbert Space to Common Sense: A Synthesis of Recent
Progress in the Interpretation of Quantum Mechanics.}

\refis{Omn7} R. Omn\`es, {\sl Rev.Mod.Phys.} {\bf 64}, 339 (1992).
{\it Consistent Intepretations of Quantum Mechanics}.

\refis{Omn8} R. Omn\`es, \pl {\bf A187}, 26 (1994).
{\it A Model for the Uniqueness of Data and Decoherent Histories.}

\refis{Omn9} R. Omn\`es. {\sl J.Stat.Phys.} {\bf 62}, 841 (1991).
{\it About the Notion of Truth in Quantum Mechanics}.


\refis{PZ} J. P. Paz and W. H. Zurek, \pr {\bf D48}, 2728 (1993).
{\it Environment-Induced Decoherence, Classicality and Consistency
of Quantum Histories.}

\refis{Sor} R. Sorkin, Syracuse preprint SU-GP-93-12-1 (1993).
{\it Quantum Mechanics as Quantum Measure Theory}.

\refis{Sau1} S. Saunders, Harvard preprint (1992).
{\it The Quantum Block Universe.}

\refis{Sau2} S. Saunders, \pl {\bf A184}, 1 (1993).
{\it Decoherence and Evolutionary Adaptation}.

\refis{Sau3} S. Saunders, Harvard preprint (1993).
{\it Decoherence, Relative States and Evolutionary Adaptation.}

\refis{Twa1} J. Twamley, Adelaide preprint ADP-93-202/M16 (1993).
{\it Inconsistency Between Alternative Approaches to Quantum
Decoherence in Special Systems.}

\refis{Twa2} J. Twamley, Adelaide preprint ADP-93-208/M19 (1993).
{\it Phase Space Decoherence: A Comparison between Consistent
Histories and Environment-Induced Superselection.}

\refis{Whe} J. A. Wheeler, in {\it Relativity, Groups and Topology},
eds. C.DeWitt and B.DeWitt (Gordon and Breach, New York, 1963).

\refis{WZ} A useful source of literature on the Copenhagen
interpretation is,
J. A. Wheeler and W. Zurek (eds.), {\it Quantum Theory and
Measurement} (Princeton University Press, Princeton, NJ, 1983).

\refis{Zur} W. Zurek, in {\it Physical Origins of Time Asymmetry}, edited by
J. J. Halliwell, J. Perez-Mercader and W. Zurek (Cambridge
University Press, Cambridge, 1994).
{\it Preferred Sets of States, Predictability, Classicality and
Environment-Induced Decoherence}.

\endreferences

\end